# Alkali and Halogen Chemistry in Volcanic Gases on Io

Laura Schaefer and Bruce Fegley, Jr.

Planetary Chemistry Laboratory, Department of Earth and Planetary Sciences
Washington University, One Brookings Dr., Campus Box 1169, St. Louis, MO 63130
E-mail: laura_s@levee.wustl.edu and bfegley@levee.wustl.edu

**ABSTRACT**

We use chemical equilibrium calculations to model the speciation of alkalis and halogens in volcanic gases emitted on Io. The calculations cover wide temperature (500-2000 K) and pressure ($10^{-6}$ to $10^{+1}$ bars) ranges, which overlap the nominal conditions at Pele (T = 1760 K, P = 0.01 bars). About 230 compounds of 11 elements (O, S, Li, Na, K, Rb, Cs, F, Cl, Br, I) are considered. The elemental abundances for O, S, Na, K, and Cl are based upon observations. CI chondritic elemental abundances relative to sulfur are used for the other alkalis and halogens (as yet unobserved on Io). We predict the major alkali species in Pele-like volcanic gases and the percentage distribution of each alkali are LiCl (73%), LiF (27%); NaCl (81%), Na (16%), NaF (3%); KCl (91%), K (5%), KF (4%); RbCl (93%), Rb (4%), RbF (3%); CsCl (92%), CsF (6%), Cs (2%). Likewise the major halogen species and the percentage distribution of each halogen are NaF (88%), KF (10%), LiF (2%); NaCl (89%), KCl (11%); NaBr (89%), KBr (10%), Br (1%); NaI (61%), I (30%), KI (9%). We predict the major halogen condensates and their condensation temperatures at P = 0.01 bar are NaF (1115 K), LiF (970 K); NaCl (1050 K), KCl (950 K); KBr (750 K), RbBr (730 K), CsBr (645 K); and sold $I_2$ (200 K). We also model disequilibrium chemistry of the alkalis and halogens in the volcanic plume. Based on this work and our prior modeling for Na, K, and Cl in a volcanic plume, we predict the major loss processes for the alkali halide gases are photolysis and/or condensation onto grains. Their estimated photochemical lifetimes range from a few minutes for alkali iodides to a few hours for alkali fluorides. Condensation is apparently the only loss process for elemental iodine. On the basis of elemental abundances and photochemical lifetimes, we recommend searching for gaseous KCl, NaF, LiF, LiCl, RbF, RbCl, CsF, and CsCl around volcanic vents during eruptions. Based on abundance considerations and observations of brown dwarfs we also recommend a search of Io's extended atmosphere and the Io plasma torus for neutral and ionized Li, Cs, Rb, and F.

**INTRODUCTION**

Several years ago we examined the chemistry of Na, K, and Cl in volcanic gases on Io and predicted that NaCl, Na, and $Na_2Cl_2$ are the major Na gases and that KCl, $K_2Cl_2$, and K are the major K gases emitted from hot volcanic vents on Io (Fegley and Zolotov 1999, 2000). Subsequently Lellouch et al. (2003) detected NaCl (g), which may originate volcanically, on Io. The purpose of the present paper is to examine the chemistry of the other alkalis (Li, Rb, Cs) and halogens (F, Cl, Br) to determine what other species of these elements may be present in Ionian volcanic gases.

Terrestrial volcanic gases and sublimates and terrestrial hot springs often have high enrichments of alkali and halogen elements (Sugiura et al. 1963, Yoshida et al. 1965, Yoshida et al. 1971, Gautier & Le Cloarec 1998). By analogy, we would also expect these elements to be present in the emissions of hot Ionian lavas. This is supported by the fact that Na, K, and Cl are observed in Io's atmosphere and in the Io plasma torus (IPT) and that NaCl (g) has been detected (Feaga et al. 2004, Küppers and Schneider 2000, Brown 2001, Lellouch et al. 2003). Furthermore, the high temperatures of some Ionian lavas are favorable for the vaporization of Na and K, as we recently showed (Schaefer and Fegley 2004a). Given the chemical similarities, we therefore expect that the less abundant alkalis (Li, Rb, and Cs), and the remaining halogens (F, Br, and I) will also be present in some volcanic gases on Io.

In this work, we extend our previous thermochemical equilibrium calculations for the system O/S/Na/K/Cl to include the less abundant alkalis (Li, Rb, and Cs) and halogens (F, Br, and I). We then examine the disequilibrium chemistry in the erupted plume. Based upon our earlier work in Moses et al. (2002), we focus on two possible loss processes for the alkali halides: photodissociation and condensation onto dust grains in the plume. Preliminary results were given by Schaefer & Fegley (2003).

**METHODS**

We modeled thermochemical equilibrium of O, S, Li, Na, K, Rb, Cs, F, Cl, Br, and I compounds as a function of temperature and pressure in a Pele-like volcanic gas. We took the nominal temperature and pressure of the Pele vent as 1760 ± 210 K and 0.01 bars based on Galileo data and our prior modeling (Lopes et al. 2001, Zolotov and Fegley 2001). The nominal elemental composition that we used is listed in Table 1. The O/S, Na/S, Cl/S, and Na/K ratios match values used in our prior work on alkali and halogen chemistry in Pele-like volcanic gases (Fegley and Zolotov 2000, Zolotov and Fegley 2001, Moses et al. 2002). As described in these papers, the nominal O, S, Na, Cl, and K abundances are based on measurements of Io's extended atmosphere and the Io plasma torus (IPT) taken from Spencer and Schneider (1996), Küppers and Schneider (2000) and Brown (2001). More recent observations by Feldman et al (2004) give a lower $Cl^{2+}/S^{2+}$ ratio (~0.021) in the IPT, while Feaga et al. (2004) report a higher Cl/S monatomic ratio (~0.10) in Io's atmosphere. Our selected Cl/S ratio of 0.045 is between these values.

Lithium, Rb, Cs, F, Br, and I have not been detected at Io to date. We estimated the abundances of these elements (which are all cosmically less abundant than Na, K, or Cl) by taking CI chondritic (i.e., proto solar) elemental ratios relative to sulfur (Lodders 2003). Figure 1 shows the abundances of the alkalis and halogens relative to sulfur in a variety of materials including: CI and LL chondrites, Juvinas (a volatile depleted basaltic achondritic (eucritic) meteorite), a continental flood basalt (BCR-1), and the bulk silicate Earth (Lodders and Fegley 1998, Raczek et al. 2001).

Based on chemical modeling of the Jovian protoplanetary subnebula by Prinn and Fegley (1981) and on cosmochemical arguments, Consolmagno (1981) and Lewis (1982) proposed that Io accreted from CM2 or CV3 carbonaceous chondrites. Based on Galileo gravity data and their interior structure modeling, Sohl et al. (2002) concluded that Io's Fe/Si mass ratio is closer to that of CM2 or CV3 chondrites than CI chondrites. In contrast Kuskov and Kronrod (2000, 2001) suggested that Io's composition is closer to that of L or LL chondrites. Both groups find that Io's Fe/Si mass ratio is less than the CI chondritic value (1.7), although the exact value is model dependent. The models suggest that the bulk composition of Io is chondritic.

The Na/S and K/S ratios in the IPT are close to the CI chondritic values while the Cl/S ratio is closer to the LL chondritic value. The sulfur normalized CI chondritic ratios for Li, Rb, Cs, F, Br, and I in Fig. 1 are probably lower limits because igneously differentiated samples such as Juvinas, the BCR-1 basalt, and the bulk silicate Earth contain less sulfur than chondritic material. However, until more data are available we use the sulfur normalized CI chondritic abundances for Li, Rb, Cs, F, Br, and I in our nominal composition. The effects of variable elemental abundances on our results will be described later in this paper.

The ideal gas thermochemical equilibrium calculations were done using the CONDOR code (Fegley and Lodders 1994) and a Gibbs energy minimization code of the type described by Van Zeggern and Storey (1970). As in our prior work we did calculations over a broad temperature and pressure range of 500-2000 K and $10^{-6}$ to $10^{+1}$ bars.

The basic assumption underlying the present work and our prior results is that volcanic gases on Io attain thermochemical equilibrium inside the volcanic conduit where the high temperatures and pressures lead to characteristic chemical reaction times that are less than the characteristic eruption time ($t_{chem} < t_{erupt}$). However, the low temperatures and pressures in the erupted volcanic plume lead to the opposite situation, namely ($t_{chem} > t_{erupt}$), and chemical equilibrium is not attained. In between these two regions, in the vicinity of the volcanic vent, $t_{chem}$ and $t_{erupt}$ become equal and quenching of the high temperature chemical equilibria occurs during the supersonic eruptions. Figure 1 of Zolotov and Fegley (1999) illustrates our basic assumption.

Zolotov and Fegley (1998a) considered the kinetics of SO destruction in Ionian volcanic gases. The SO abundance decreases as the volcanic gas cools because SO is converted back to sulfur vapor and $SO_2$ via net thermochemical reactions such as

$$4 \, SO = 2 \, SO_2 + S_2 \qquad (1)$$

Reaction (1) proceeds via a series of elementary reactions. The elementary reaction

$$SO + SO \rightarrow SO_2 + S \qquad (2)$$

is important for SO destruction in photochemical models (e.g., Summers and Strobel 1996) and is plausibly the rate determining step for SO destruction in volcanic gases. The characteristic time ($t_{chem}$) for SO destruction via reaction (2) is given by

$$t_{chem}(SO) = 1/(k_2[SO]) \qquad (3)$$

where the square brackets denote number density and the rate constant $k_2$ (Summers and Strobel 1996) is taken as

$$k_2 = 5.8 \times 10^{-12} \exp(-1760/T) \text{ cm}^3 \text{ s}^{-1}. \qquad (4)$$

Table 1 of Zolotov and Fegley (1998a) shows that $t_{chem}$ values for SO destruction in cooling volcanic gases are always less than typical eruption times of 150-250 seconds over a range of plausible vent temperatures (1000-1400 K) and pressures ($10^{-3}$ to $10^{-5}$ bars). For example, at 1400 K and $10^{-3}$ bars, $t_{chem}$ is 0.008 seconds. Decreasing the pressure to $10^{-5}$ bars increases $t_{chem}$ to 0.2 seconds, which is still much less than the typical eruption time. The latter are computed from literature values of 0.2 km s$^{-1}$ for the eruption velocity (Kieffer 1984) and 30-50 km depth of volcanic source regions (Spencer and Schneider 1996). The conclusion that $t_{chem} < t_{erupt}$ in the volcanic conduit is robust and is unchanged if other plausible values for eruption velocities and source region depths are used instead. However it is also important to demonstrate that reactions involving alkalis and halogens rapidly reach chemical equilibrium in Ionian volcanic gases. We consider several reactions to make this point.

Our prior work (Fegley and Zolotov 2000) and our results below show that $Na_2Cl_2$ gas becomes more abundant with decreasing temperature. It forms by the elementary reaction

$$NaCl + NaCl \rightarrow Na_2Cl_2 \qquad (5)$$

with a rate constant $k_5 = 1.33 \times 10^{-10}$ cm$^3$ s$^{-1}$ (Mallard et al. 1998). The corresponding chemical lifetime for NaCl (g) over the 1400-2000 K range is

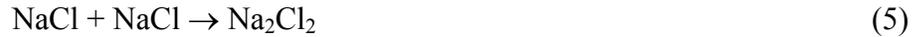
$$t_{chem}(NaCl) = 1/(k_5[NaCl]) \sim 4.7 \times 10^{-6} \text{ s}. \qquad (6)$$

Sodium chloride is the major Cl-bearing gas and one reaction for its formation is

$$Na + Cl_2 \rightarrow NaCl + Cl \qquad (7)$$

which has a rate constant $k_7 = 7.3 \times 10^{-10}$ cm$^3$ s$^{-1}$ (Mallard et al. 1998). The corresponding chemical lifetime for $Cl_2$ (g) is

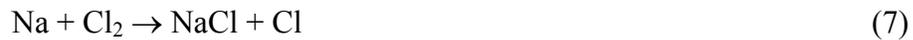
$$t_{chem}(Cl_2) = 1/(k_7[NaCl]) \sim 4.0 \times 10^{-6} \text{ s}. \qquad (8)$$

Finally, we calculate $t_{chem}$ for an exchange reaction between NaCl and KCl gases that was in our model for photochemistry of volcanic plumes on Io (Moses et al. 2002)

$$K + NaCl \rightarrow KCl + Na. \qquad (9)$$

Our estimated rate constant $k_9 = 10^{-14}$ cm$^3$ s$^{-1}$ (Moses et al. 2002). The corresponding chemical lifetime for NaCl (g) is

$$t_{chem} (NaCl) = 1/(k_9[K]) \sim 11 \text{ s}. \qquad (10)$$

Thus it appears volcanic gases on Io, like those on Earth (Symonds et al. 1994), are in chemical equilibrium when they are erupted. The agreement between the predicted and observed abundances of SO in volcanic gases on Io (Zolotov and Fegley 1998a, McGrath et al. 2000, dePater et al. 2002) also indicates that the gases are chemically equilibrated. We explicitly assume this is so in this work.

**RESULTS**

Table 2 and Figs. 2-5 show the major species of S, O, Li, Na, K, Rb, Cs, F, Cl, Br, and I in our model volcanic gas. Figures 2 and 3 show the abundances of volcanic gas and condensates at 1760 K (the nominal temperature at Pele) as a function of the total pressure. The shaded area in these figures shows the pressure range that we previously calculated for Pele (Zolotov and Fegley 2001). Figures 4 and 5 are complementary and display the percentage distribution of these eleven elements between volcanic gas and condensates at 0.01 bars (the nominal total pressure at Pele) as a function of temperature from 500 to 2000 K. The shaded area shows the temperature range (1760 ± 210 K) reported for Pele by Lopes et al. (2001). Table 3 lists the condensates and their condensation temperatures at 0.01 bars pressure.

We discuss these results below. We first describe chemistry at the nominal temperature and pressure for the Pele vent and then we discuss the effects of variable pressure and temperature. Several generalizations can be made. First, the equilibrium chemistry of the more abundant alkali metals (Na and K) and halogens (Cl and F) controls that of the less abundant elements (Li, Rb, Cs, Br, I) to a large extent. This control gives abundance curves with complex shapes as a function of temperature and pressure for many of the less abundant species shown in the figures. Second, the abundances of atoms increase with decreasing pressure at the expense of molecular gases, e.g., the monatomic alkalis (Fig. 2) and monatomic halogens (Fig. 3) increase at the expense of alkali halides. Molecules with weaker bonds (i.e., smaller dissociation energies) are more strongly affected than those with stronger bonds (larger dissociation energies). This point is illustrated by comparing the abundance curves for NaF, NaCl, NaBr, and NaI in Fig. 3a-d. The abundance of NaF is essentially constant down to $10^{-6}$ bars, that of NaCl decreases by about a factor of three from $10^{-4}$ to $10^{-6}$ bars, the abundance of NaBr drops by more than a factor of 10 between $10^{-3}$ and $10^{-6}$ bars, and the abundance of NaI decreases by about a factor of 1,000 from $10^{-1}$ to $10^{-6}$ bars. Third, thermal ionization becomes increasingly important with decreasing pressure and the abundances of the singly ionized alkalis and electrons increase with decreasing pressure (Fig. 2). Finally, the abundances of atoms and ions increase with increasing temperature at the expense of molecular and monatomic gases. For example, CsI decomposes to its constituent atoms and Cs is ionized to $Cs^+$ as shown in Fig. 4e.

*Sulfur and Oxygen*: Figures 3e and 5e present our results for sulfur and oxygen. These results agree with our prior work in which we discuss the effects of temperature, pressure, and bulk composition on sulfur and oxygen chemistry of Ionian volcanic gases (Fegley and Zolotov 2000, Zolotov and Fegley 1998a, b, 1999, 2000, 2001). Briefly, $SO_2$ is the major gas at all temperatures and pressures studied. At the nominal temperature and pressure for Pele, the next most important gases are $S_2$, SO, $S_2O$, S, and $S_3$ in order of decreasing abundance. With decreasing temperature about 0.4% of total sulfur condenses as sodium sulfate ($Na_2SO_4$) at 1350 K and 0.2% of all sulfur condenses as potassium sulfate ($K_2SO_4$) at 620 K. The $S_3$ to $S_8$ allotropes also become more abundant at the expense of $S_2$ with decreasing temperature, and elemental sulfur condenses below 500 K.

*Sodium and Potassium:* We predict that the major sodium gases emitted by Pele are NaCl (81%), Na (16%), and NaF (3%). The major potassium gases are KCl (91%), K (5%), and KF (4%). These results agree with our previous conclusions that the major Na and K gases are the chloride and the monatomic gas, respectively (Fegley and Zolotov 2000). The chloride is the most abundant gas for each element at high pressures (see Figs. 2a-b). As pressure drops, the abundance of the chloride decreases whereas the abundance of the monatomic gas increases. The third most abundant gas for each element is the fluoride, except at the highest pressures where liquid sodium sulfate ($Na_2SO_4$) becomes stable and dimeric sodium chloride gas ($Na_2Cl_2$) becomes more abundant than NaF. The abundance of NaF is essentially independent of the total pressure, while that of KF increases slightly with decreasing pressure.

The results in Fig. 4a-b predict that all sodium and potassium are in the gas at the temperatures observed for Pele. As temperature decreases below 1400 K, sodium and potassium both form several different condensates. With decreasing temperature sodium condenses as thenardite ($Na_2SO_4$) at 1345 K, as villiaumite (NaF) at 1115 K, and as halite (NaCl) at 1050 K. At 500 K, halite makes up ~ 93% of total sodium, while $Na_2SO_4$(c) and NaF(c) make up ~3.5% each. Potassium condenses as sylvite (KCl) at 950 K and as KBr (c) at 750 K. At 600 K, sylvite converts to arcanite ($K_2SO_4$). At 500 K, $K_2SO_4$ (c) makes up ~99.6 % of total K, and KBr (c) ~0.4% of total potassium.

*Lithium:* We predict that the two most important Li gases at Pele are LiCl (~73%) and LiF (~27%). Less than 1% of all lithium is in Li, LiBr, and LiO (see Fig. 2c). With decreasing pressure, the abundance of LiCl drops, whereas the abundance of LiF increases, so LiF is more abundant below $10^{-6}$ bars. Monatomic Li is the third most abundant gas, and it is 1.5-2.5 orders of magnitude less abundant than the Li halides. The abundance of monatomic Li decreases with increasing pressure. The fourth most abundant gas is LiBr (Fig. 2c), which has an approximately constant abundance from $10^{+1}$ to $10^{-3}$ bars. Its abundance then decreases more than an order of magnitude between $10^{-3}$ and $10^{-6}$ bars.

Lithium chemistry at 0.01 bars as a function of temperature is shown in Figure 4c. The abundance of LiF decreases slightly and the abundances of LiCl and monatomic Li increase from 1110 to 2000 K. Lithium fluoride gas has its peak abundance at 1115 K where solid NaF condenses. Below this temperature, LiF (g) decreases while LiCl, LiBr, and $Li_2F_2$ increase in abundance. The abundances of all gases drop sharply once solid LiF condenses at ~940 K. All lithium is condensed as LiF by 500 K.

*Rubidium:* We predict that the major Rb gases at Pele are RbCl (93%), monatomic Rb (4%), and RbF (3%). Figure 2d shows that this sequence remains the same over the range of pressures calculated for Pele. Monatomic Rb and $Rb^+$ become more abundant with decreasing pressure. Conversely, Rb becomes less abundant than RbF as pressure increases above ~4 bars. Rubidium bromide and other Rb-bearing gases are always less abundant than RbCl, Rb, and RbF.

Figure 4d shows Rb chemistry as a function of temperature. The key point is that RbCl, Rb, and RbF remain the major Rb gases over the range of temperatures reported for Pele. However the abundances of these gases decrease at lower temperatures: Rb (~1350 K), RbF (~1100 K), and RbCl (~800 K). Rubidium bromide and iodide have fairly constant abundances from 2000 to 1100 K. Their abundances increase at temperatures below 1100 K. Solid RbBr condenses at 730 K, and all rubidium is found in RbBr (c) by 500 K.

*Cesium:* As shown in Fig. 2e, we predict that the major Cs gases at Pele are CsCl (92%), CsF (6%), and monatomic Cs (2%). However, $Cs^+$ becomes increasingly important with decreasing pressure. It is the third most abundant Cs gas from $10^{-3.5}$ to $10^{-5}$ bars, and the second most abundant gas at lower pressures.

Figure 4e shows Cs chemistry as a function of temperature. Cesium chloride, CsF, and Cs are the three major gases down to about 1200 K where CsBr becomes as abundant as Cs. Cesium bromide becomes increasingly important with decreasing temperature. Solid CsBr condenses at 645 K and at 500 K all cesium is found in CsBr (c).

*Chlorine:* Figures 3b and 5b illustrate chlorine chemistry. We predict that the major chlorine gases at Pele are NaCl (89%) and KCl (11%) in agreement with our previous work (Fegley and Zolotov 2000). The next most abundant alkali chloride gas is LiCl, which is ~2 orders of magnitude less abundant than KCl (g). Monatomic Cl and $Cl^-$ become more abundant at lower pressures than at Pele but the $Cl/Cl^-$ ratio is over 100. Conversely polymeric alkali chlorides, in particular $Na_2Cl_2$, become more abundant at higher pressures than at Pele.

Figure 5b shows all of the major chlorine gases at a constant pressure as a function of temperature. Sodium chloride and KCl are the major gases down to about 1200 K where $Na_2Cl_2$ becomes more abundant than KCl. Chlorine condenses as halite (NaCl) at ~ 1050 K and as sylvite (KCl) at ~950 K. Sylvite remains stable down to 620 K where it is converted into arcanite ($K_2SO_4$), with the chlorine forming more halite. At 500 K, all chlorine is found in halite.

*Fluorine:* Figure 3a illustrates fluorine chemistry as a function of pressure. We predict that the major fluorine gases at Pele are NaF (88%), KF (10%), and LiF (2%). This sequence is independent of total pressure down to $10^{-6}$ bars where monatomic F becomes as abundant as LiF. Figure 5a shows fluorine chemistry as a function of temperature. Sodium fluoride, KF, and LiF are the major gases down to about 1280 K where $Na_2F_2$ becomes more abundant than LiF. The abundance of $Na_2F_2$ increases with decreasing temperature until it is the second most abundant fluorine gas at 1115 K where NaF (villiaumite) condenses. Fluorine gas chemistry is especially complicated in this temperature range with a number of polymeric alkali fluoride species being important. Solid LiF condenses at ~970 K and removes the remaining fluorine from the gas. At 500 K, 93% of all fluorine is in villiaumite with the remaining 7% in LiF.

*Bromine*: As shown in Fig. 3c, we predict that the major bromine gases at Pele are NaBr (89%), KBr (10%), and Br (1%). The Br gas abundance increases with decreasing pressure. It is the third most abundant gas from about $10^{-1}$ to $10^{-3}$ bars, the second most abundant gas from $10^{-3}$ to $10^{-4}$ bars, and the most abundant gas from $10^{-4}$ to $10^{-6}$ bars. Lithium bromide becomes the third most abundant Br gas at pressures greater than $10^{-1}$ bars, but still accounts for less than 1% of total bromine at these pressures.

Figure 5c shows bromine chemistry as a function of temperature. Sodium bromide and KBr are the two major gases over a wide temperature range. The third most abundant gas switches from Br to LiBr to RbBr with decreasing temperature. Bromine chemistry is especially complex in the 500-1000 K range due to formation of dimeric alkali bromides and the sequential condensation of KBr (750 K), RbBr (730 K), and CsBr (645 K). At 500 K, RbBr makes up ~58% of total bromine, KBr ~39% and CsBr ~3%.

*Iodine*: Figure 3d illustrates iodine chemistry as a function of pressure. We predict that the major iodine gases at Pele are NaI (61%), monatomic I (30%), and KI (9%). The abundance of I increases and those of NaI and KI decrease with decreasing pressure until essentially all iodine is present as monatomic I gas. All other iodine gases are insignificant at 1760 K, but RbI, CsI, and dimeric alkali iodides become increasingly important at lower temperatures as shown in Fig. 5d.

Iodine chemistry, like that of Br, is complicated at lower temperatures. Potassium, rubidium and cesium iodides all increase in abundance below 1000 K and peak in abundance at ~950K, 730 K and 650 K, respectively. These peak abundances are correlated with the condensation temperatures of KBr, RbBr, and CsBr. However, none of the alkali iodides condenses, and their abundances decrease at lower temperatures. Monatomic iodine peaks in abundance at ~650 K, and is then converted into $I_2$ (g). Iodine finally condenses as elemental iodine ($I_2$) at ~200 K.

The high volatility of iodine may seem surprising, but it is also characteristic of iodine geochemistry on Earth where iodine is mobilized as methyl iodide ($CH_3I$) and $I_2$ vapor (Muramatsu and Wedepohl 1998, Honda 1970). Honda (1970) showed that the thermochemical reaction

$$2H_2O + 2I_2 = 4\,HI + O_2 \tag{11}$$ 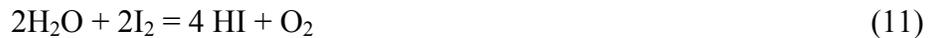

controls the $HI/I_2$ ratio in terrestrial volcanic gases. Terrestrial volcanic gases are generally steam-rich (e.g., see Table 6.13 in Lodders and Fegley 1998), and reaction (11) yields large $HI/I_2$ ratios in most terrestrial volcanic gases. However, Io is bone dry and there is no steam to react with $I_2$ vapor.

In contrast to bromine, iodine is not enriched in terrestrial volcanic sublimates, but apparently remains in the gas (Honda 1970). Our calculations of volatile element geochemistry on Venus showed similar behavior with all iodine in the gas as HI or elemental iodine (Schaefer and Fegley 2004b).

*Silicon Tetrafluoride*: Silicon tetrafluoride ($SiF_4$) is observed in volcanic gases from Aso, Etna, Iwodake, Popocatépetl, and Vulcano (Francis et al. 1996; Love et al. 1998; Mori et al. 2002). The $SiF_4$ is detected by observing its 9.7 μm band with infrared (IR) absorption spectroscopy, which is also used to measure $SO_2/SiF_4$ molar ratios. The $SO_2/SiF_4$ ratios vary from 71 at Iwodake to over 3000 at Aso. Mori et al. (2002) studied emissions from Iwodake in some detail and derived an average $SiF_4/HF$ molar ratio of 0.57 from their IR measurements of $SO_2$, $SiF_4$, HCl, and HF.

On Earth, $SiF_4$ is formed by reaction of HF with silica

$$4\,HF\,(g) + SiO_2\,(silica) = SiF_4\,(g) + 2\,H_2O\,(g) \tag{12}$$ 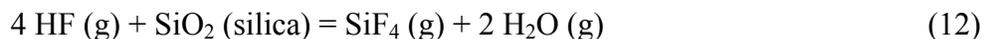

Io is bone-dry and HF gas should not be present. However alkali fluoride gases may react with pure silica to give $SiF_4$ via reactions analogous to equation (12). We therefore included pure silica in another set of computations that were done as a function of temperature at the nominal pressure of ~0.01 bars for Pele and also as a function of pressure at the nominal temperature of 1760 K. We found $SiF_4$ is only important at low temperatures (see Fig. 5f), and that essentially no $SiF_4$ is produced at 1760 K. A comparison of Fig. 5f with Fig. 5a shows that fluorine chemistry is basically the same down to 1115 K where solid NaF condenses. However, there are several differences at lower temperatures. Sodium fluoride disappears at about 820 K instead of remaining stable down to 500 K. Also, $SiF_4$ becomes much more abundant below 1115 K and is the major fluorine-bearing gas below about 900 K.

Table 2 summarizes our predictions for the major alkali- and halogen-bearing gases emitted in Pele-like volcanic gases on Io. All alkalis and halogens are present as gases at 1760 K and 0.01 bars, the nominal conditions for Pele. The major condensates that form at lower temperatures are listed in Table 3.

**DISCUSSION**

Once the alkali halides are volcanically erupted into Io's tenuous atmosphere, what happens to them? We therefore decided to look at the photochemistry of the alkali halides to estimate how long the gases will survive once they are released into Io's atmosphere. We then make recommendations on what gases to search for based upon abundance and estimated lifetime.

*Photochemistry of the Alkali Halides*

Previous work by Moses et al. (2002) examined the photochemistry of Na, K and Cl species at Io. Their work suggests that the main loss processes for NaCl (g) and KCl (g) are photochemical destruction and condensation onto dust grains inside volcanic plumes via the reactions:

$$XCl + h\upsilon = X + Cl \qquad (13)$$
$$XCl + dust = XCl\ (condensed) \qquad (14)$$

where X = Na or K. The photochemical lifetimes in the steady state model of Moses et al. (2002) for NaCl (g) and KCl (g) are 1.8 hours each. Here we consider similar reactions for the other alkali halides. We estimated the photodissociation rate, $J_1$ (at zero optical depth) from the equation:

$$J_1 = \Sigma \sigma_n I_n \qquad (15)$$

where $\sigma_n$ is the absorption cross section in cm$^2$ and $I_n$ is the solar flux in cm$^{-2}$ s$^{-1}$ nm$^{-1}$, for the reactions:

$$XZ + h\upsilon = X + Z \qquad (16)$$

where X represents the alkalis and Z represents the halogens. The calculated $J_1$ values are listed in Table 4 with the corresponding chemical lifetimes. We used the solar flux outside Earth's atmosphere from Warneck (2000) scaled to the radial distance of Jupiter and the absorption cross sections for the alkali chlorides, bromides and iodides measured by Davidovits & Brodhead (1967). We could not find absorption cross sections for alkali fluoride gases in the literature so we estimated their chemical lifetimes from trends in the $J_1$ values and bond energies for the chlorides, bromides, and iodides in Table 4.

Other determinations of the absorption cross sections for alkali halides have been made for NaCl (g) at room temperature by Silver et al. (1986), CsI (g) at room temperature by Grossman et al. (1977), and the Cs halides at extremely high temperatures (1900-2000 K) by Mandl (1971) and Mandl (1977). Moses et al. (2002) calculated a $J_1$ value for NaCl (g) of 8.8×10$^{-5}$ s$^{-1}$ from the room temperature cross section data of Silver et al. (1986). This value is almost 100 times smaller than our calculated value due to differences in the absorption cross sections of Silver et al. (1986) and Davidovits & Brodhead (1967). Davidovits and Brodhead (1967) measured cross sections at higher temperatures (~1000-1100 K) than Silver et al. (1986); the higher vibrational populations associated with higher temperatures and the possible presence of dimers account for their larger cross sections. We therefore suggest that our calculated $t_{chem}$ values are lower limits on the actual lifetimes of the alkali halides. Also, the absorption of UV light by species other than alkali halides will lower the $J_1$ values and therefore increase the photochemical lifetimes of the alkali halides.

We also looked at the heats of reaction for chemical reactions of the type:

$$XM + Z = XZ + M \qquad (17)$$

where XM is an alkali halide and Z is an abundant element such as O, S, Na, K, or Cl, in order to qualitatively assess whether these reactions are important loss processes for the alkali halides. Most reactions of this type are endothermic, and therefore unlikely to be important loss processes on Io. Those that may be important are mentioned below.

*Fluorides*: Very little data is available on the photochemistry of alkali fluorides. Mandl (1971) determined the photodissociation cross section of CsF (g) at ~2000 K. Compared with the photodissociation cross sections of CsBr, CsCl, and CsI at similar temperatures (Mandl 1977), the cross section of CsF is smaller. This makes sense from comparisons of bond strengths and the $J_1$ values for the other halides, which show similar trends. With the exception of CsBr, which does not follow the trend, the $J_1$ values and bond energies decrease from chloride to bromide to iodide. We therefore expect that the photochemical lifetimes of the alkali fluorides should be significantly longer than for the corresponding chlorides, bromides and iodides. The values for the chemical lifetimes of the alkali fluorides given in Table 4 were extrapolated from the values for the other halides and are considered lower limits on the actual lifetimes. Given the long photodissociation lifetimes (1.2 - 11 hours) and assuming that condensation of alkali fluorides proceeds at rates similar to the NaCl condensation rate estimated by Moses et al. (2002), we suggest that condensation is plausibly more important for the loss of alkali fluorides than photodissociation.

*Chlorides*: The major loss processes of NaCl (g) and KCl (g) have been studied in detail previously, as discussed above. Based upon similar $J_1$ values, we suggest that RbCl (g) and CsCl (g) will behave in much the same way. The major loss processes of these gases would therefore be photodissociation and condensation, respectively. In the case of LiCl (g), the photodissociation lifetime is much longer (~3.6 hrs) and is more similar to the lifetimes of the alkali fluorides than of the alkali chlorides. We therefore suggest that condensation may be a more important loss process for LiCl (g) than photodissociation.

*Bromides and Iodides*: The major loss processes for the alkali bromides will plausibly be photodissociation and condensation. The chemical lifetimes calculated from the $J_1$ values for photodissociation are less than half an hour for all of the alkali bromides except LiBr (g). Since these are lower limits on the photochemical lifetime for the reasons discussed above, these species may have sufficiently long enough lifetimes to begin condensing on dust grains. The major loss process for alkali iodides is plausibly photodissociation, since our calculated $J_1$ values for the reactions: $XI + h\nu = X + I$ all give photochemical lifetimes of less than ten minutes. These species should be almost instantaneously lost from the volcanic plumes. Given their short lifetimes and low abundances, alkali iodides will probably not be observable at Io. Instead, we expect iodine to occur primarily as I (g) and $I_2$ (g). We calculated a $J_1$ value for molecular iodine from absorption cross section data from Myer and Samson (1970) which gives a photochemical lifetime of several days. The exchange reactions:

$$XM + Cl = XCl + M \tag{18}$$

where X represents one of the alkalis and M is either Br or I, are highly exothermic, and may also be important loss processes for the alkali bromides and iodides.

*Variable elemental abundances*

As mentioned earlier, Li, Rb, Cs, F, Br, and I have not been detected at Io to date and their abundances were estimated by taking CI chondritic (i.e., solar) elemental ratios relative to solar (Lodders 2003). Their estimated abundances would be about the same if we chose CI chondritic ratios relative to Na, Cl, or K because the observed IPT elemental abundances for S, Na, Cl, and K are similar to the CI chondritic values (see Fig. 1). However, there is uncertainty in the observed IPT elemental abundances since the IPT is both spatially and temporally variable (Spencer and Schneider 1996). Observations show that sulfur and oxygen chemistry is different at different volcanic hot spots (e.g., McGrath et al. 2000), and as discussed earlier (Fegley and Zolotov 2000), there are good reasons for differences in the alkali and halogen chemistry at different hot spots. Thus our model may not be applicable to all hot spots on Io.

The formation and interior structure models discussed earlier indicate that Io may have a bulk chondritic composition. However, as noted by Gerlind Dreibus in her referee report "Figure 1 illustrates perfectly, that in the case of igneously differentiated bodies like Vesta and Earth the loss of halogens is higher compared to the loss of alkali elements." Io is apparently differentiated with a metallic core and silicate lithosphere (e.g., Sohl et al. 2002). Thus bulk alkali to halogen ratios on Io could be smaller than the CI chondritic values used here.

Additionally, even if bulk Io has a chondritic composition (CI or CM2 or CV or L or LL chondritic), there are a variety of mechanisms that will fractionate the alkalis and halogens with different efficiencies, including planetary differentiation (core vs. silicate), rock/melt partitioning of magma sources, and vapor/melt partitioning of molten lavas.

To simplify matters, we assume that we are computing the equilibrium gas chemistry above a lava of CI chondritic composition. However, we still then have to consider vapor/melt partitioning. A survey of the literature on terrestrial volcanic gases indicates that degassing efficiencies for different elements are dependent upon a number of factors including temperature, pressure, lava composition, oxidation state, $H_2O$ and S contents, etc. However, general trends for the relative degassing efficiencies of the alkalis and halogens seem to be Cs~Li~Rb>K>Na and I>Cl~Br>F (see for example Gautier & Le Cloarec 1998, Yoshida et al. 1965, Taran et al. 1995, Symonds et al. 1987, Symonds et al. 1990, Lambert et al. 1988, Crowe et al. 1987, and Chukarov et al. 2000). The degassing efficiency of chlorine is sometimes greater and sometimes less than that of bromine depending upon lava composition and temperature, whereas fluid/melt partition coefficients indicate that the degassing efficiencies for the halogens are I>Br>Cl (Bureau et al. 2000). We would therefore expect that X/Na (where X = Li, Cs, and Rb) should be larger in a gas than in the lava from which it arises. For the halogens, the volcanic gas data indicates that I/Cl in the gas should be larger than in the lava, Br/Cl should remain approximately constant, and F/Cl in the gas should be smaller than in the lava. Since we are taking CI chondritic abundances of the minor alkalis and halogens with respect to sulfur and not sodium or chlorine, our model ratios for Z/Cl (where Z = F, Br, and I) are significantly smaller than the CI chondritic values and our model ratios for X/Na are larger than the CI chondritic values. Our model is therefore qualitatively consistent with the degassing efficiencies for the alkalis and for F if we assume a CI chondritic lava composition. While our values for I/Cl and Br/Cl are possibly smaller than would be advised from the volcanic gas data, these elements are much less abundant than F and Cl (see Table I) and should therefore not have much of an effect on the alkali chemistry, since the major gases of the alkalis and halogens are consistent over a wide range of abundances. It is not until the total abundance of halogens is greater than that of the alkalis ($\Sigma$ (halogens) > $\Sigma$ (alkalis)) that a major change in the gas speciation occurs, as was discussed in our previous modeling (Fegley and Zolotov 2000). The total abundance of halogens may exceed that of alkalis at low temperature hot spots, such as Loki or Prometheus. In these cases sulfur chlorides, sulfur oxychlorides, and other sulfur halogen compounds may be formed as illustrated in Figs. 3d and 5d of Fegley and Zolotov (2000). We focus here on the case where Σ (alkalis) > Σ (halogens) that probably applies to Pele at T = 1760 K.

*Recommended species for observational searches*

Based upon solar system elemental abundances, our thermochemical equilibrium calculations for volcanic gases, and our photochemical models of alkali halide destruction in Io's atmosphere, we suggest that the best candidates for observations in order of decreasing abundance are, respectively: KCl, NaF, KF, LiCl, LiF, RbCl, RbF, CsCl, and CsF. Potassium chloride, NaF, and KF are all relatively abundant and relatively long-lived. Potassium chloride, which is the most abundant gas we predict that has not already been detected at Io, has previously been observed in the circumstellar envelope of a cool star (Cernicharo et al. 2000, Cernicharo & Guelin 1987). The Li, Rb, and Cs fluorides and LiCl are much less abundant than their sodium and potassium counterparts but are predicted to be relatively long-lived. Lines of monatomic Li, Rb, and Cs are observed in the optical and near IR spectra of stars and the latter two alkalis are observed in brown dwarfs (Lodders 1999). The Li, Rb, and Cs fluorides have bands in the far IR and mm region, but we are unaware of any detections in the ISM, circumstellar envelopes of cool stars, or other astronomical environments. Rubidium and Cs chlorides have shorter lifetimes and their detection may only be possible during on-going volcanic eruptions. The alkali bromides may not be detectable at Io, given their lower abundances and shorter lifetimes. We do not expect alkali iodides to be detectable at Io. Iodine should occur as $I_2$ (g) and I (g).

In the Io plasma torus (IPT), we would expect F to be the next most abundant halogen (after Cl), followed by Li and Br, based upon solar system abundances. By analogy with Cl and considering the similarities in ionization potentials, we would expect F to occur in the IPT mainly as F II and F III. Cesium, rubidium and lithium may also be observable in the IPT with most Cs and Rb present as ions while Li may be distributed between monatomic and ionized Li. While they are not as abundant as F, the neutrals of these elements—Cs I, Rb I, and Li I—have previously been observed in the atmospheres of brown dwarfs at wavelengths of 8521/8943 Å, 7800/7948 Å, and 6708 Å, respectively (e.g., Burgasser et al. 2003, Griffith and Yelle 2000, Lodders 1999, Salim et al. 2003). Thus monatomic Cs, Rb, and Li may also be observable at Io.

## SUMMARY

We calculated the equilibrium distribution of the alkali and halogen elements in a volcanic gas at the temperature and pressure for the Pele vent. We predict that the major species of the alkalis at these conditions are the chlorides, fluorides and monatomic gases. The major species of the halides are the sodium and potassium halides and the monatomic gases. Our results for Na, K, and Cl confirm our previous conclusions for the chemistry of these elements (Fegley and Zolotov 2000). The effect of the other alkalis and halogens upon the chemistry of Na, K, and Cl is minor. We predict a variety of condensates, including $Na_2SO_4$, NaCl, NaF, $K_2SO_4$, KCl, KBr, LiF, RbBr, and CsBr, which may be present near volcanic vents on Io. Iodine does not condense until ~ 200 K as elemental iodine. We also estimated the photochemical lifetimes of the alkali halide gases in Io's atmosphere. We find that fluorides are more stable than chlorides followed by bromides and then iodides. We therefore recommend searching for the following gases based upon abundances and photochemical lifetimes: KCl, NaF, KF, LiCl, LiF, RbCl, RbF, CsCl, and CsF. Finally based on our modeling we also recommend searches for the elements Li, Rb, Cs, and F in Io's extended atmosphere and in the IPT in both neutral and ionized forms.

## ACKNOWLEDGMENTS

This work was supported by Grant NAG5-11958 from the NASA Planetary Atmospheres Program. We thank Katharina Lodders for helpful discussions, and Michael Summers and Gerlind Dreibus for constructive reviews.

| Table 1. Nominal Abundance Model | | |
|:---:|:---:|:---:|
| Element | Abundance | CI Abundance |
| O | 1.521 | 16.97 |
| S | 1.000 | 1.000 |
| Na | 0.050 | 0.129 |
| Cl | 0.045 | 0.012 |
| K | 0.005 | $8.05\times10^{-3}$ |
| F | $1.89\times10^{-3}$ | $1.89\times10^{-3}$ |
| Li | $1.25\times10^{-4}$ | $1.25\times10^{-4}$ |
| Br | $2.54\times10^{-5}$ | $2.54\times10^{-5}$ |
| Rb | $1.48\times10^{-5}$ | $1.48\times10^{-5}$ |
| I | $2.24\times10^{-6}$ | $2.24\times10^{-6}$ |
| Cs | $8.25\times10^{-7}$ | $8.25\times10^{-7}$ |

| Table 2. Major Gas Species at 1760 K and $10^{-2}$ bars | | |
|---|---|---|
| Element | Major gases | % of Element |
| Na | NaCl | 81 % |
|  | Na | 16 % |
|  | NaF | 3 % |
| K | KCl | 91 % |
|  | K | 5% |
|  | KF | 4% |
| Li | LiCl | 73 % |
|  | LiF | 27 % |
| Rb | RbCl | 93 % |
|  | Rb | 4 % |
|  | RbF | 3 % |
| Cs | CsCl | 92 % |
|  | CsF | 6 % |
|  | Cs | 2 % |
| F | NaF | 88 % |
|  | KF | 10 % |
|  | LiF | 2 % |
| Cl | NaCl | 89 % |
|  | KCl | 11 % |
| Br | NaBr | 89 % |
|  | KBr | 10 % |
|  | Br | 1 % |
| I | NaI | 61 % |
|  | I | 30 % |
|  | KI | 9 % |

| Table 3. Major Condensates at $10^{-2}$ bars ||| 
| Element | Condensates | Temperature (K) |
| --- | --- | --- |
| Na | $Na_2SO_4$ | ~1350 |
|    | NaF | ~1115 |
|    | NaCl | ~1050 |
| K  | KCl | ~950 - 620 |
|    | $K_2SO_4$ | ~620 |
|    | KBr | ~750 |
| Li | LiF | ~970 |
| Rb | RbBr | ~730 |
| Cs | CsBr | ~645 |
| F  | NaF | ~1115 |
|    | LiF | ~970 |
| Cl | NaCl | ~1050 |
|    | KCl | ~950 |
| Br | KBr | ~750 |
|    | RbBr | ~730 |
|    | CsBr | ~645 |
| I  | I | ~200 |

**Table 4.** Photochemical Lifetimes of the Alkali Halides

| Species | $D°_{298}$ (kJ mol$^{-1}$)[a] | $J_1$ (s$^{-1}$) | $t_{chem}$ (min.) |
|---|---|---|---|
| LiF | 577±21 | — | ~686[b] |
| LiCl | 469±13 | 7.72×10$^{-5}$ | 216 |
| LiBr | 419±4 | 3.72×10$^{-4}$ | 45 |
| LiI | 345±4 | 2.69×10$^{-3}$ | 6.2 |
| | | | |
| NaF | 519 | — | ~71[b] |
| NaCl | 412±8 | 6.46×10$^{-4}$ | 26 |
| NaBr | 367±1 | 2.00×10$^{-3}$ | 8.3 |
| NaI | 304±2 | 7.30×10$^{-3}$ | 2.3 |
| | | | |
| KF | 498±3 | — | ~226[b] |
| KCl | 433±8 | 2.67×10$^{-4}$ | 62 |
| KBr | 380±1 | 1.13×10$^{-3}$ | 15 |
| KI | 325±1 | 5.20×10$^{-3}$ | 3.2 |
| | | | |
| RbF | 494±21 | — | ~177[b] |
| RbCl | 428±8 | 3.40×10$^{-4}$ | 49 |
| RbBr | 381±4 | 1.46×10$^{-3}$ | 11 |
| RbI | 319±2 | 8.54×10$^{-3}$ | 2.0 |
| | | | |
| CsF | 519±8 | — | ~232[b] |
| CsCl | 448±8 | 2.17×10$^{-4}$ | 77 |
| CsBr | 389±4 | 1.97×10$^{-3}$ | 8.5 |
| CsI | 337±2 | 1.96×10$^{-3}$ | 8.5 |

[a] $D°_{298}$ values are taken from the CRC Handbook of Chemistry and Physics
[b] $t_{chem}$ values for all alkali fluorides are estimates. See text for details.

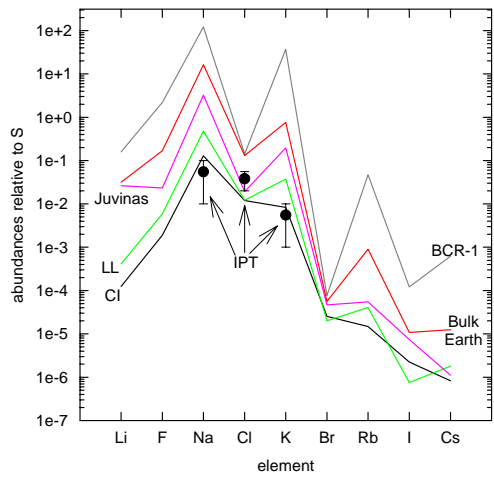

Figure 1. A comparison of alkali and halogen abundances with respect to sulfur in CI chondrites, LL chondrites, a eucrite achondrite meteorite (Juvinas), Columbia River flood basalt (BCR-1) and in the bulk silicate Earth. Data are from Lodders and Fegley (1998).

Figure 2. Equilibrium chemistry of the alkalis at the nominal temperature of 1760 K for Pele as a function of total pressure: (a) sodium, (b) potassium, (c) lithium, (d) rubidium, and (e) cesium. The shaded region corresponds to the calculated range ($10^{-1.53}$ to $10^{-2.21}$ bars) and nominal pressure ($10^{-1.96}$ bars) for Pele at 1760 K (Zolotov and Fegley 2001).

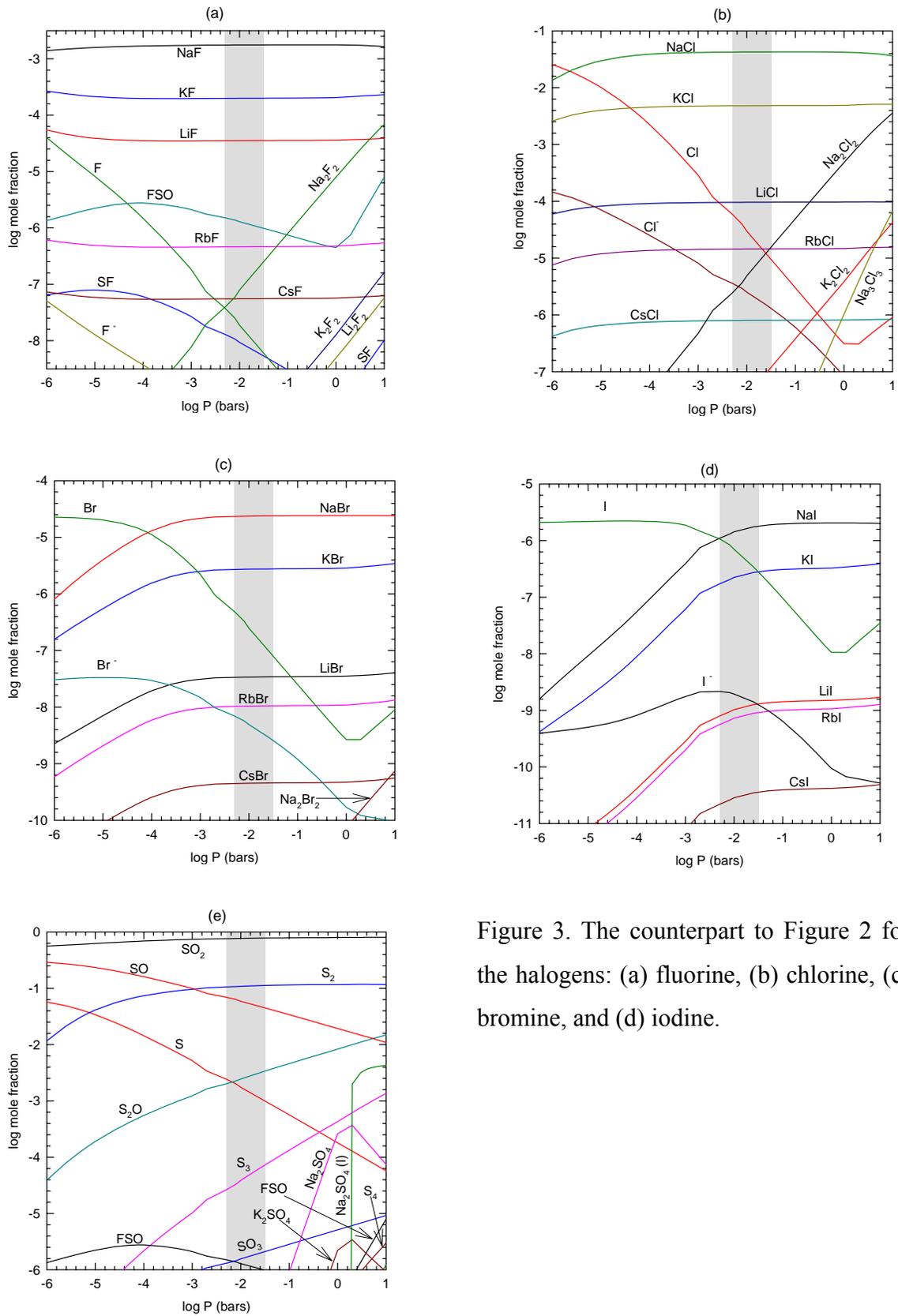

Figure 3. The counterpart to Figure 2 for the halogens: (a) fluorine, (b) chlorine, (c) bromine, and (d) iodine.

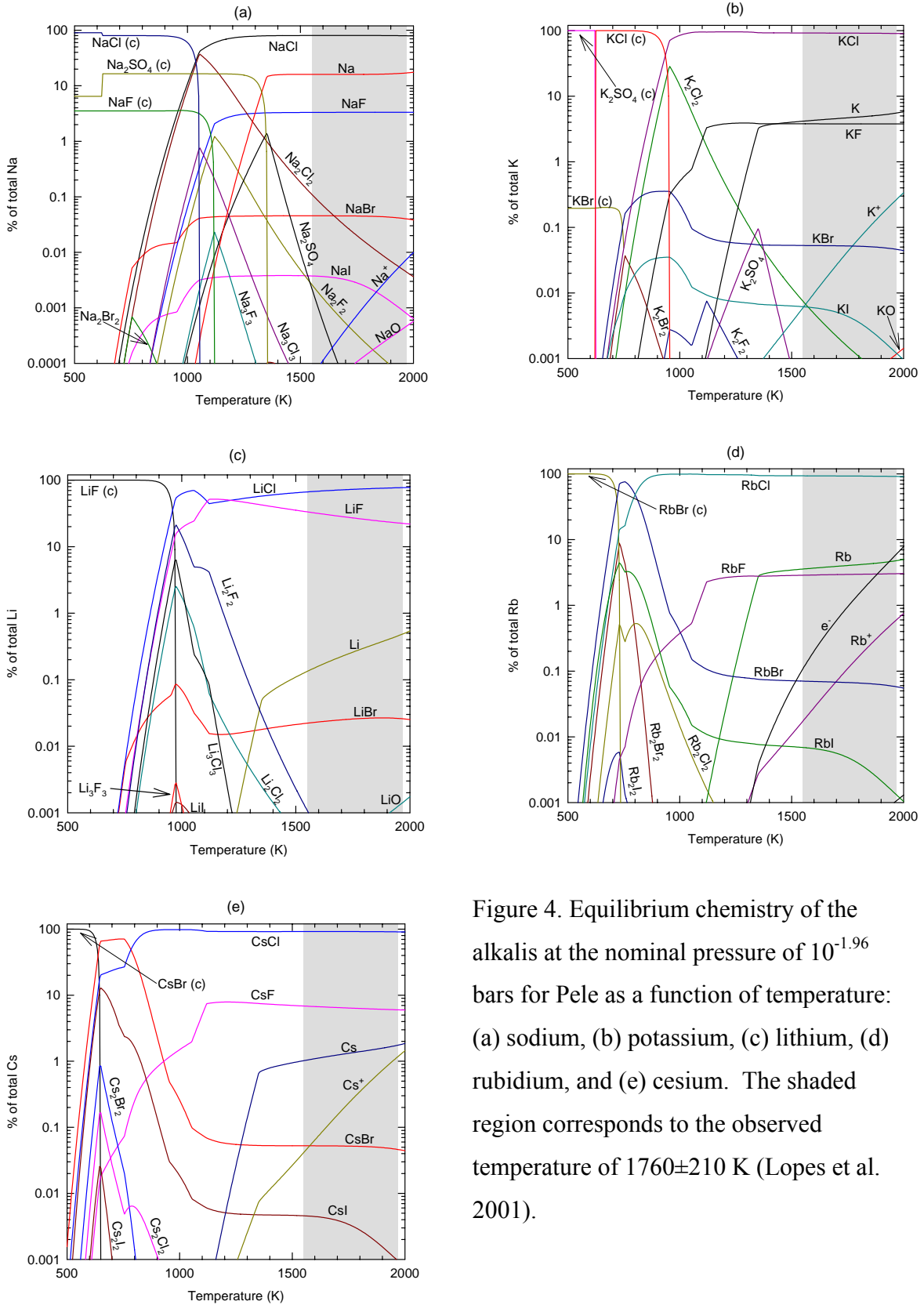

Figure 4. Equilibrium chemistry of the alkalis at the nominal pressure of $10^{-1.96}$ bars for Pele as a function of temperature: (a) sodium, (b) potassium, (c) lithium, (d) rubidium, and (e) cesium. The shaded region corresponds to the observed temperature of $1760\pm210$ K (Lopes et al. 2001).

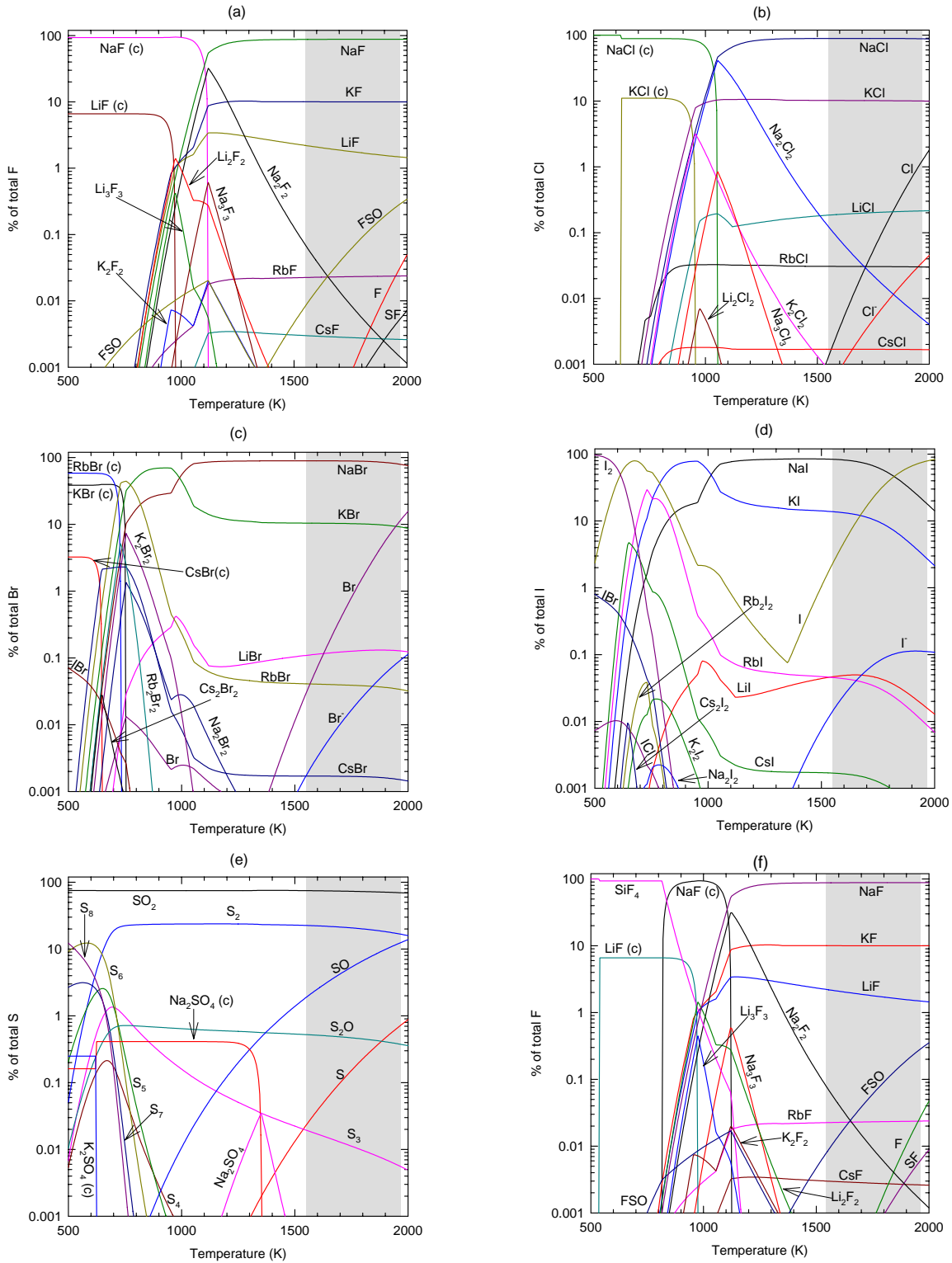

Figure 5. The counterpart to Figure 4 for the halogens: (a) fluorine, (b) chlorine, (c) bromine, (d) iodine, and (e) fluorine chemistry including silicon tetrafluoride.